\documentclass[a4paper]{jpconf}
\usepackage{graphicx}

\def\bea{\begin{eqnarray}}
\def\eea{\end{eqnarray}}

\def\pp{\mbox{$p$-$p$} }
\def\auau{\mbox{Au-Au} }

\def\pbpb{\mbox{Pb-Pb} }
\def\aa{\mbox{A-A} }
\def\nn{\mbox{N-N} }

\begin{document}
\title{The ``soft ridge'' -- is it initial-state geometry or
modified jets?}

\author{Thomas A.~Trainor}

\address{CENPA 354290, University of Washington, Seattle WA, USA}

\begin{abstract}
A same-side (SS, on azimuth $\phi$) 2D peak in measured angular correlations from 200 GeV \pp collisions exhibits properties expected for jet formation. In more-central \auau collisions the SS peak becomes elongated on pseudorapidity $\eta$ and the transverse momentum $p_t$ structure is modified. In the latter case the SS 2D peak has been referred to as a ``Soft Ridge,'' and arguments have been presented that the elongated peak represents flow phenomena (``triangular'' and ``higher harmonic'' flows), possibly related to the initial-state \aa geometry. In this presentation I demonstrate that ``higher harmonic flows'' are related to SS 2D peak properties and review evidence for a jet interpretation of the SS peak for all \auau centralities.
\end{abstract}

\section{Introduction -- what is the same-side 2D peak?}

Minimum-bias ($p_t$-integral) 2D angular correlations can be plotted on difference variables $\eta_\Delta = \eta_1 - \eta_2$ (pseudorapidity) and $\phi_\Delta = \phi_1 - \phi_2$ (azimuth). For all collision systems a same-side (SS, $|\phi_\Delta| < \pi/2$) 2D peak is observed. A 1D peak on azimuth is observed within the complementary away-side (AS) interval along with a quadrupole component $\cos(2\phi_\Delta)$ unrelated to the two peaks. Interpretation of the SS 2D peak mechanism is currently hotly debated.

 \begin{figure}[h] \hfil
  \includegraphics[width=1.5in,height=1.4in]{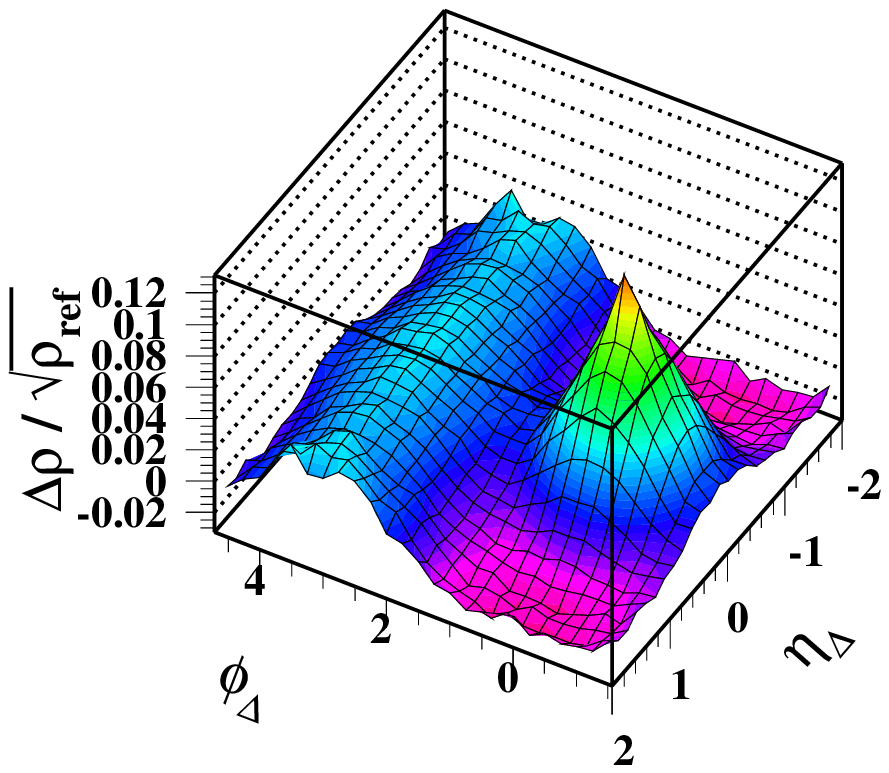} 
\put(-100,75) {\bf(a)}
 \includegraphics[width=1.5in,height=1.4in]{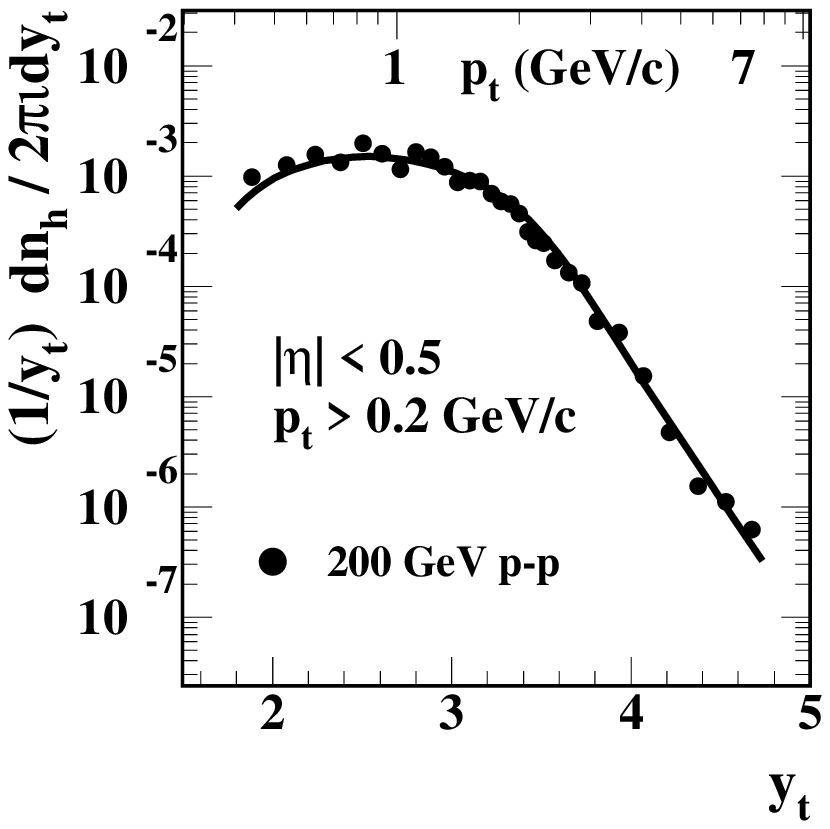}
 \put(-23,75) {\bf (b)}
 \includegraphics[width=1.5in,height=1.4in]{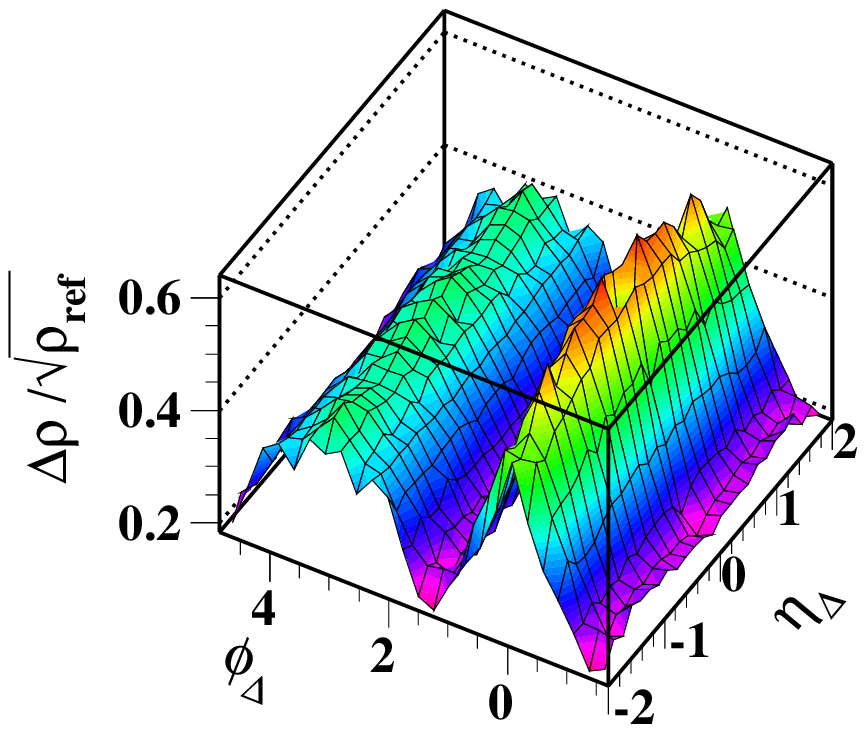}
\put(-100,75) {\bf (c)}
  \includegraphics[width=1.5in,height=1.4in]{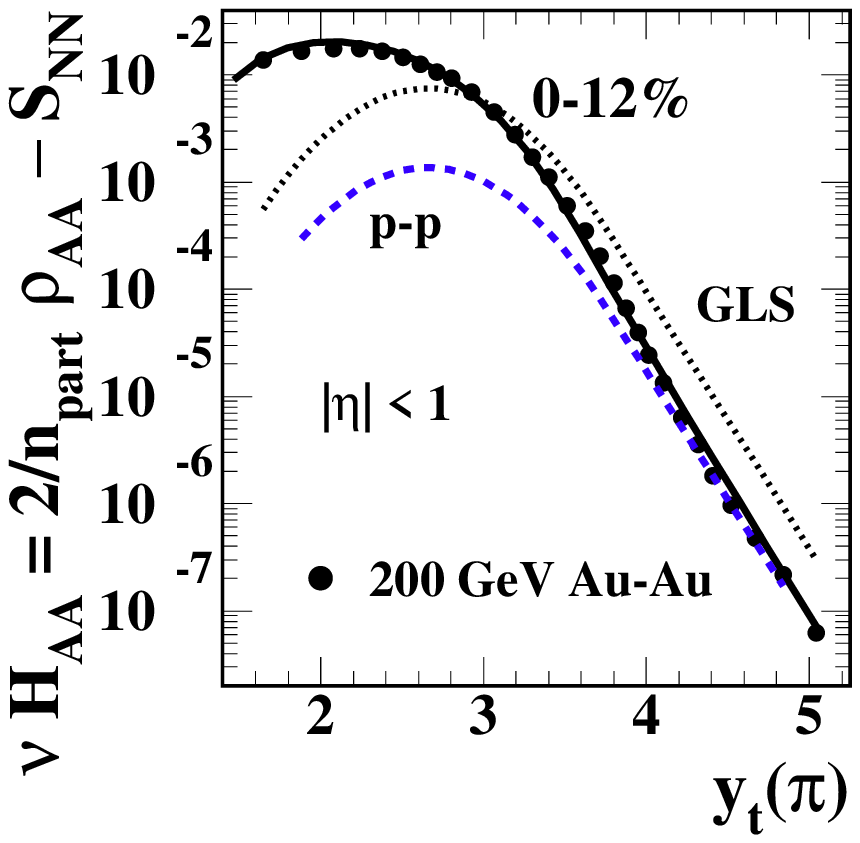} 
 \put(-23,75) {\bf (d)}
\caption{\label{tfig1}
(a) Angular-correlation hard component (jets), 200 GeV p-p, (b) p-p spectrum hard component (points) with pQCD description (curve) (c) jet correlations, 0-5\% central Au-Au, (d) Au-Au spectrum hard component (points) with pQCD description (solid curve).
}  
 \end{figure}

Figure~\ref{tfig1} (a) shows 2D angular correlations from \pp collisions for $p_t > 0.5$ GeV/c (hard component)~\cite{porter2,porter3}. The SS 2D peak is elongated 2:1 on $\phi$. The corresponding spectrum hard component in panel (b) (points from Ref.~\cite{ppprd}) is described by a pQCD calculation (solid curve) down to 0.5 GeV/c ($y_t \approx 2$, $y_t \equiv \ln[(m_t + p_t)/m_\pi]$), strongly supporting a jet interpretation~\cite{ppprd,fragevo}. Figure~\ref{tfig1} (c) shows correlations from 0-5\% central Au-Au collisions. The SS peak is strongly elongated 1:3 on $\eta$, but the corresponding spectrum hard component (points from Ref.~\cite{hardspec}) in panel (d) is still described by a pQCD calculation (solid curve), albeit with modified fragmentation~\cite{fragevo}. The dotted curve represents a GLS reference (Glauber linear superposition, \aa transparency) for the most-central data ($\nu \approx 6$). The dashed curve represents \pp data in panel (b) ($\nu \approx 1.25$). Fragmentation modification is controlled by a single parameter determined at 10 GeV/c ($y_t = 5$) relative to GLS. Panel (d) can be compared with plots of ratio $R_{AA}$ in which critical jet-related spectrum structure at lower $p_t$ is obscured~\cite{hardspec,nohydro}.

Persistence of jets in more-central \auau collisions is inconvenient for the conventional ``perfect liquid'' interpretation of RHIC data. The response consists of describing the SS 2D peak as a ``soft ridge'' attributed to flow mechanisms~\cite{glasma,glasma2}. Jets are converted to flows by imposition of Fourier analysis: 
(a) fit 1D projections onto $\phi_\Delta$ with a Fourier series,
(b) interpret each series term as a ``harmonic flow,''
(c) attribute flows to conjectured A-A initial-state geometry~\cite{isgeom}.
Despite such efforts detailed analysis confirms the jet interpretation and falsifies flow conjectures.

\section{Jet phenomenology}

The SS 2D peak must be interpreted in the context of the {\em full centrality evolution} of two-particle correlations. Centrality is measured by mean participant pathlength $\nu = 2N_{bin} / N_{part} \in [1,6]$.

 \begin{figure}[h] \hfil
  \includegraphics[width=1.5in,height=1.4in]{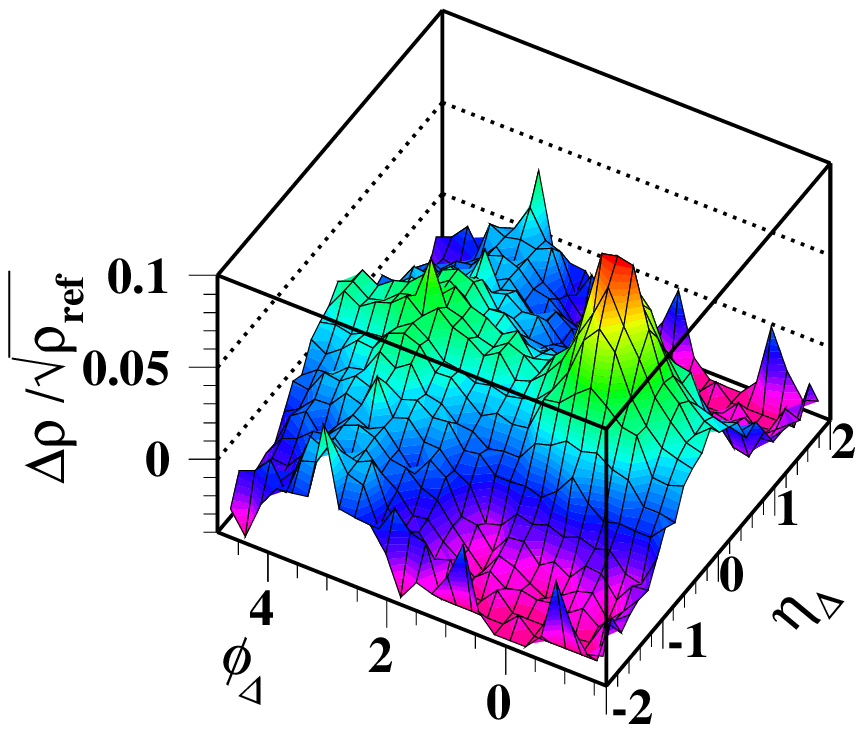}
 \put(-99,75) {\bf(a)}
 \includegraphics[width=1.5in,height=1.4in]{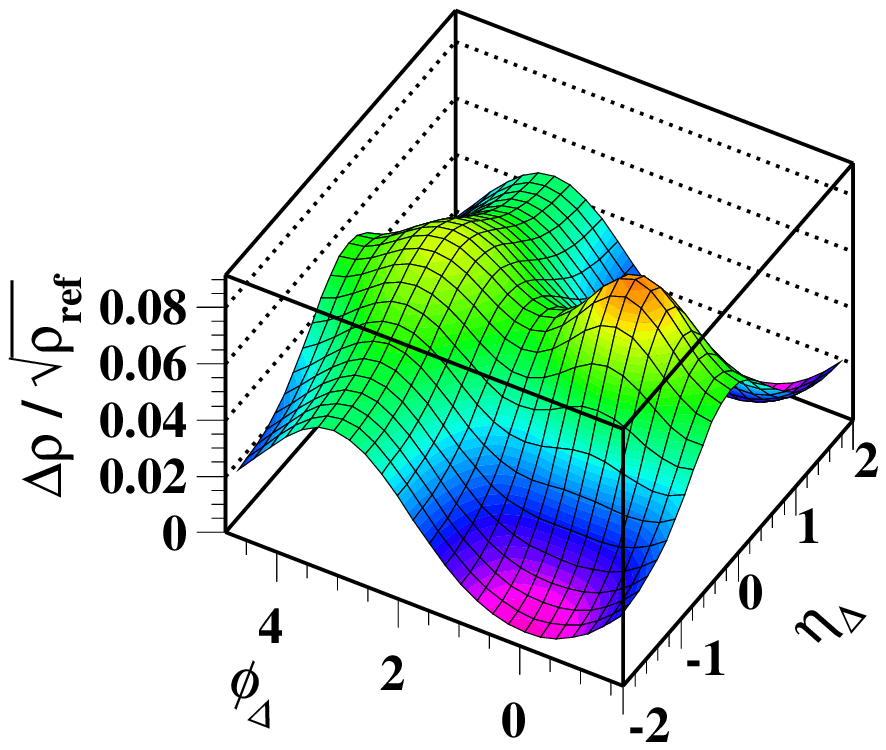}
 \put(-99,75) {\bf(b)}
  \includegraphics[width=1.5in,height=1.4in]{sextupole200-10-4}
 \put(-99,75) {\bf(c)}
  \includegraphics[width=1.5in,height=1.4in]{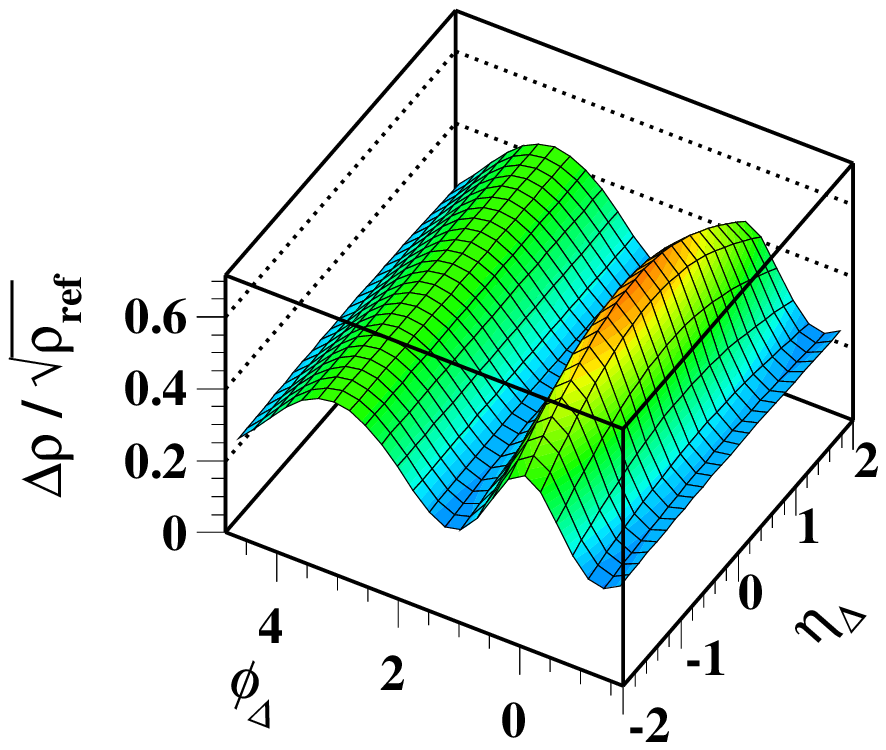}
 \put(-99,75) {\bf(d)}
\hfil
\caption{\label{tfig2} Angular correlations from 200 GeV Au-Au collisions for
(a) 85-95\% data,  (b) 2D model fit, (c) 0-5\% data, (d) 2D model fit.
}  
 \end{figure}

Figure~\ref{tfig2} shows $p_t$-integral 2D angular correlation data for (a) 85-95\% central (approximately \nn collisions, $\nu \approx 1.25$) and (c) 0-5\% central ($\nu \approx 6$) 200 GeV \auau collisions~\cite{anomalous}. The data are modeled accurately by the same multi-element 2D model function for all centralities, as in panels (b) and (d) which indicate the absolute jet structure relative to a fitted zero.

 \begin{figure}[h] \hfil
  \includegraphics[width=1.5in,height=1.37in]{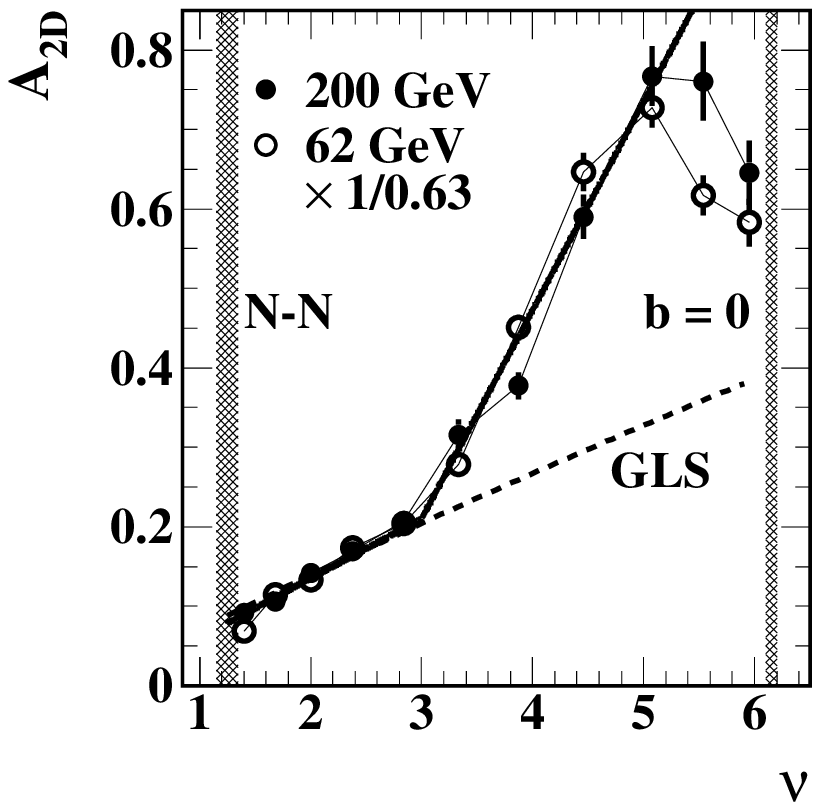}
 \put(-25,22) {\bf(a)}
  \includegraphics[width=1.5in,height=1.4in]{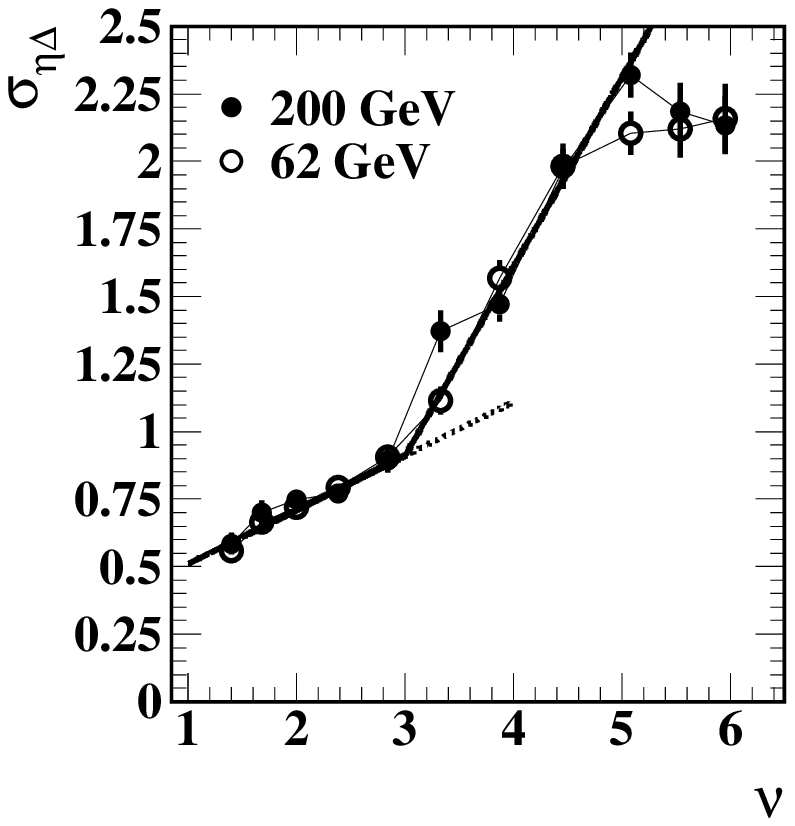}
 \put(-25,22) {\bf(b)}
  \includegraphics[width=1.5in,height=1.4in]{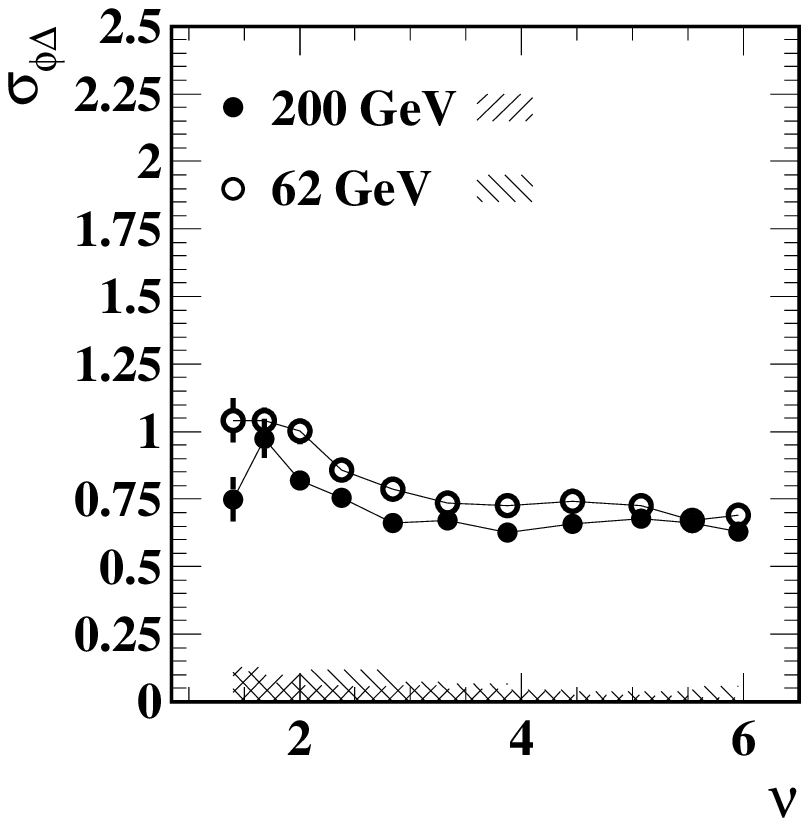}
 \put(-25,22) {\bf(c)}
  \includegraphics[width=1.5in,height=1.4in]{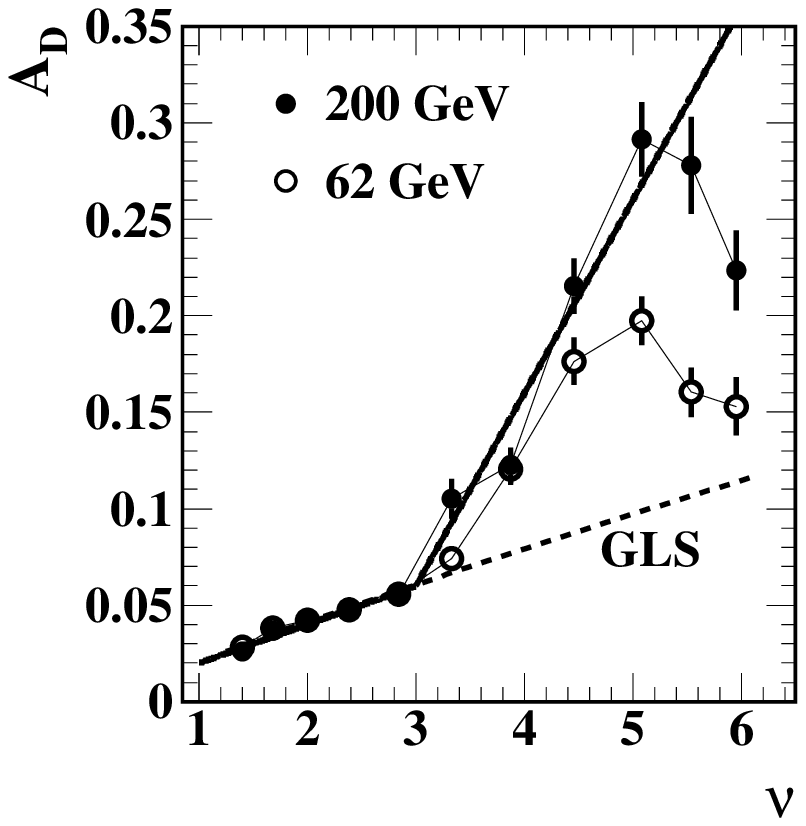}
 \put(-25,22) {\bf(d)}
\hfil
\caption{\label{tfig3} 2D model-fit parameters from Au-Au collisions for (a) same-side (SS) 2D peak amplitude, (b) SS peak $\eta$ width, (c) SS peak $\phi$ width,  (d) away-side (AS) 1D peak amplitude.
}  
 \end{figure}

Figure~\ref{tfig3} shows individual fit parameters for SS 2D [(a) amplitude, (b) $\eta$ width and (c) $\phi$ width] and AS 1D [(d) amplitude] peaks, with full coverage of \auau centrality from \nn to central \auau collisions. The SS and AS peak amplitudes are simply proportional for all centralities (within data uncertainties), consistent with scattered-parton (dijet) momentum conservation. The observed SS 2D peak is monolithic and simply described by a 2D Gaussian for all centralities; no separate ``ridge'' is resolved. Bold lines are added in three panels to emphasize the {\em sharp transition} (ST) in slopes at $\nu \approx 3$ (corresponding to $\sigma(b) / \sigma_0 \approx 0.5$). Below the ST (more-peripheral collisions) the peak properties follow Glauber linear superposition (GLS) trends expected for jets formed in {\em transparent} \aa collisions (no medium).

\section{Minijets and hadron production}

By interpreting SS peak properties in Fig.~\ref{tfig3} in terms of a jet mechanism and calculating a pQCD jet frequency  we can predict the fragment yields corresponding to measured 2D jet correlations.

 \begin{figure}[h] \hfil
  \includegraphics[width=1.5in,height=1.37in]{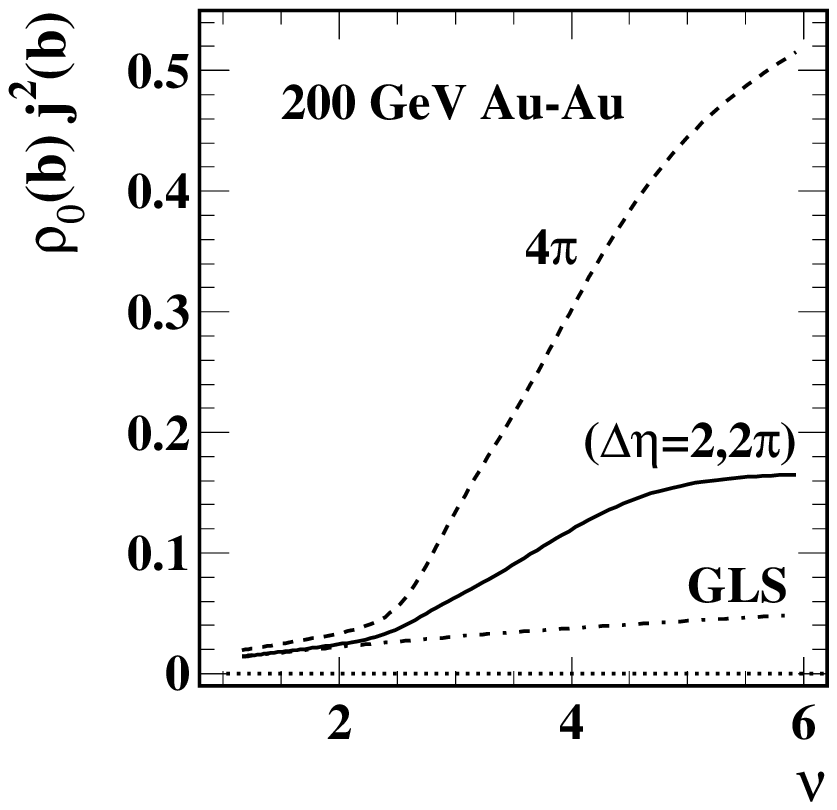}
  \put(-21,70) {\bf(a)}
 \includegraphics[width=1.5in,height=1.4in]{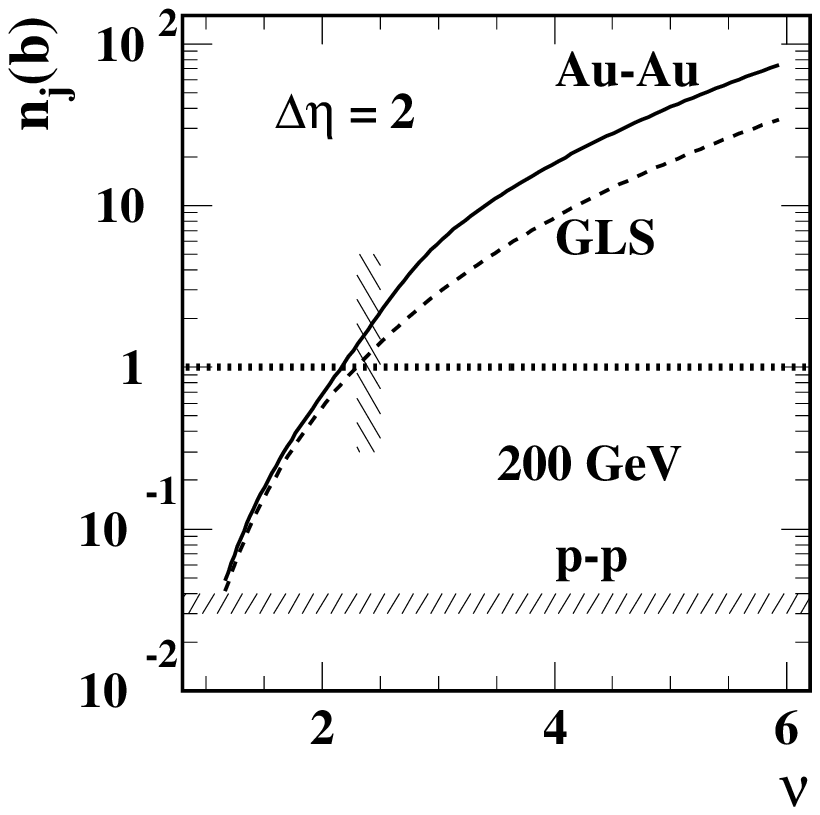}
   \put(-21,70) {\bf(b)}
 \includegraphics[width=1.5in,height=1.4in]{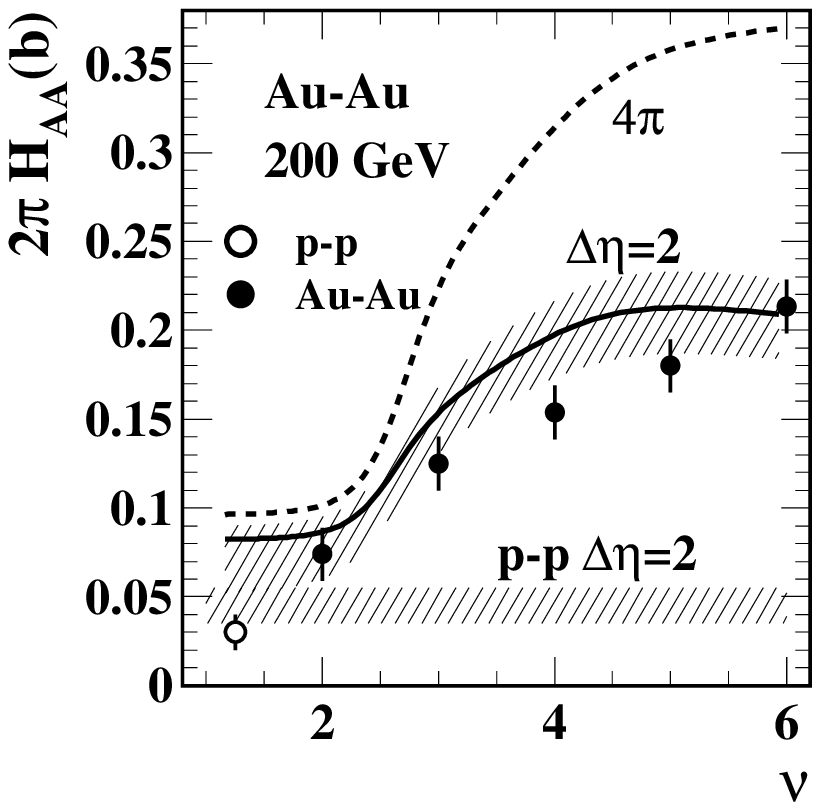}
  \put(-21,77) {\bf(c)}
  \includegraphics[width=1.5in,height=1.4in]{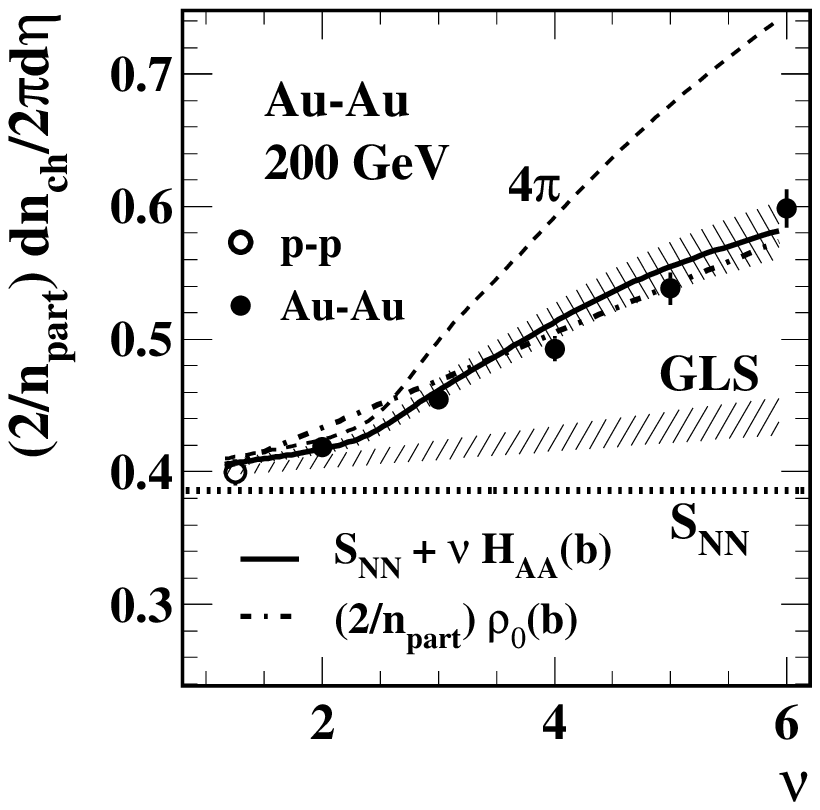}
  \put(-21,80) {\bf(d)}
\hfil
\caption{\label{tfig4} Centrality ($\nu$) distribution of
(a) SS 2D peak volume, (b) jet number within $\Delta \eta = 2$, (c) spectrum hard component, (d) predicted total hadron yield (solid curve) vs data (points).
}  
 \end{figure}

Figure~\ref{tfig4} (a) shows the jet-related pairs per particle  $\rho_0(b) j^2(b)$ averaged over the angular acceptance, where $\rho_0$ is the single-particle 2D angular density and $j^2$ is the SS-peak volume (as a pair ratio) averaged over the acceptance~\cite{jetspec}. Panel (b) shows the pQCD-predicted mean jet number per event in angular acceptance $\Delta \eta = 2$ for 200 GeV \auau collisions~\cite{fragevo,jetspec}. Combining the two solid curves we obtain the predicted 1D angular density (on $\eta$) of jet fragments plotted as the spectrum hard component $2\pi H_{AA}$ (solid curve) in panel (c) compared to measured spectrum hard components from Ref.~\cite{hardspec}. Panel (d) shows the hard component from Panel (c) combined with a per-participant-pair soft-component 2D density 0.38 held fixed for all \auau centralities.

This analysis establishes a quantitative relation among (a) jet-related 2D angular correlations, (b) jet-related spectrum hard components and (c) pQCD-calculated dijet frequencies, strongly supporting a jet interpretation of the SS 2D peak for all \auau centralities. This description of hadron production at RHIC agrees with spectrum data within uncertainties  and implies that 1/3 of final-state hadrons in central \auau collisions are contained within resolved jets.

\section{The nonjet azimuth quadrupole $\bf v_2\{2D\}$}

The third significant feature of 2D angular correlations is the {\em nonjet} azimuth quadrupole represented by $A_Q\{2D\}(b) = \rho_0(b) v_2^2\{2D\}(b)$, where $\{2D\}$ signifies 2D model fits to data.

 \begin{figure}[h] \hfil
  \includegraphics[width=1.9in,height=1.6in]{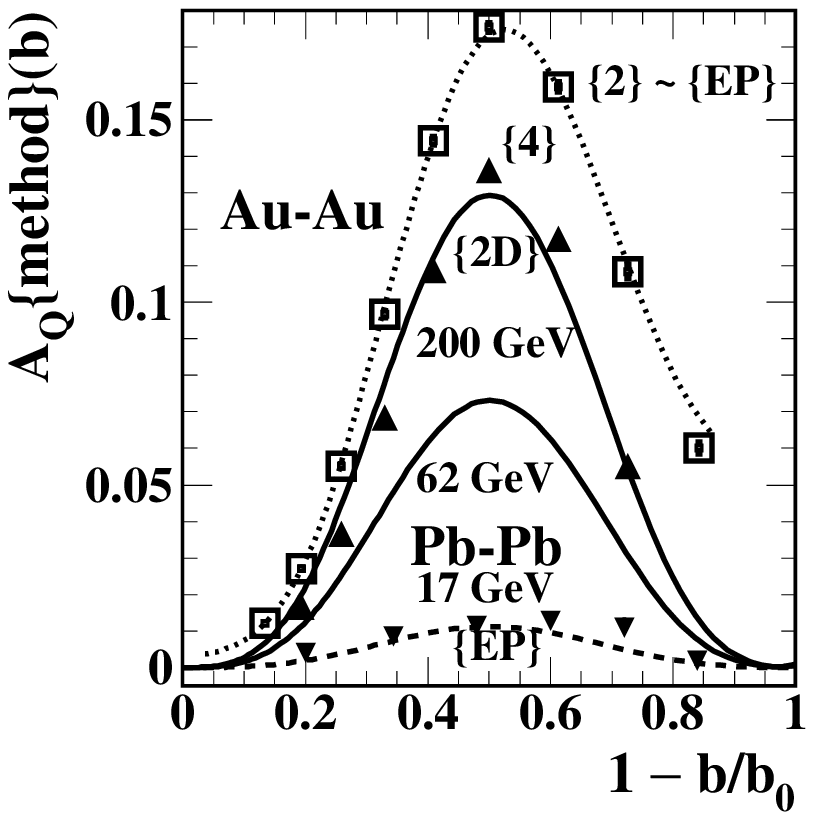}
  \put(-99,65) {\bf(a)}  \hfil
  \includegraphics[width=1.9in,height=1.6in]{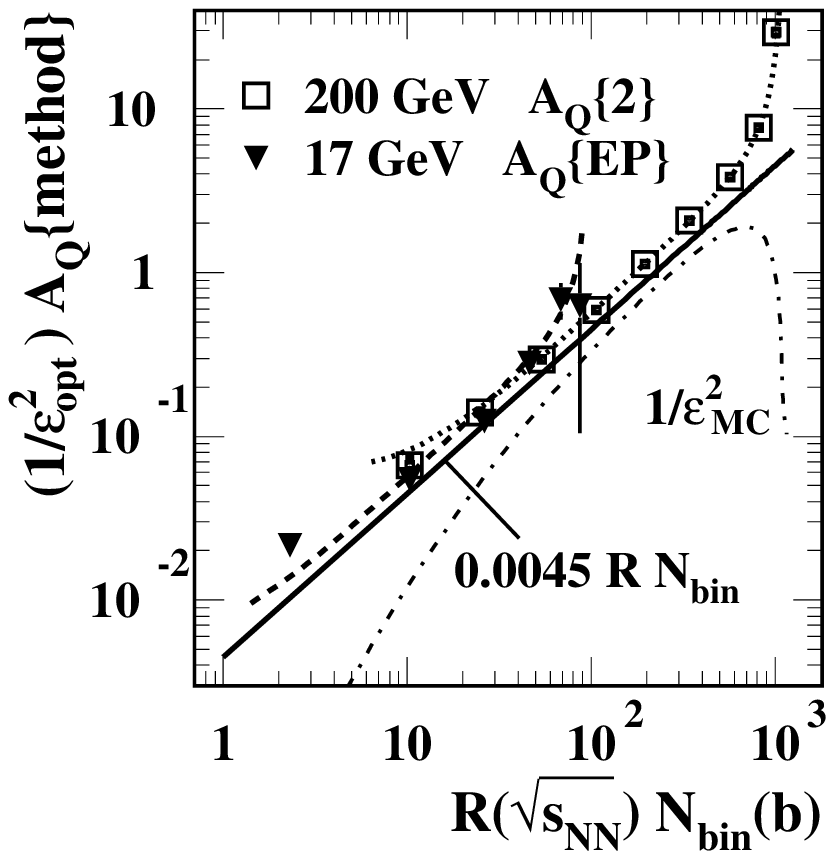}
   \put(-99,65) {\bf(b)} \hfil
 \includegraphics[width=1.9in,height=1.6in]{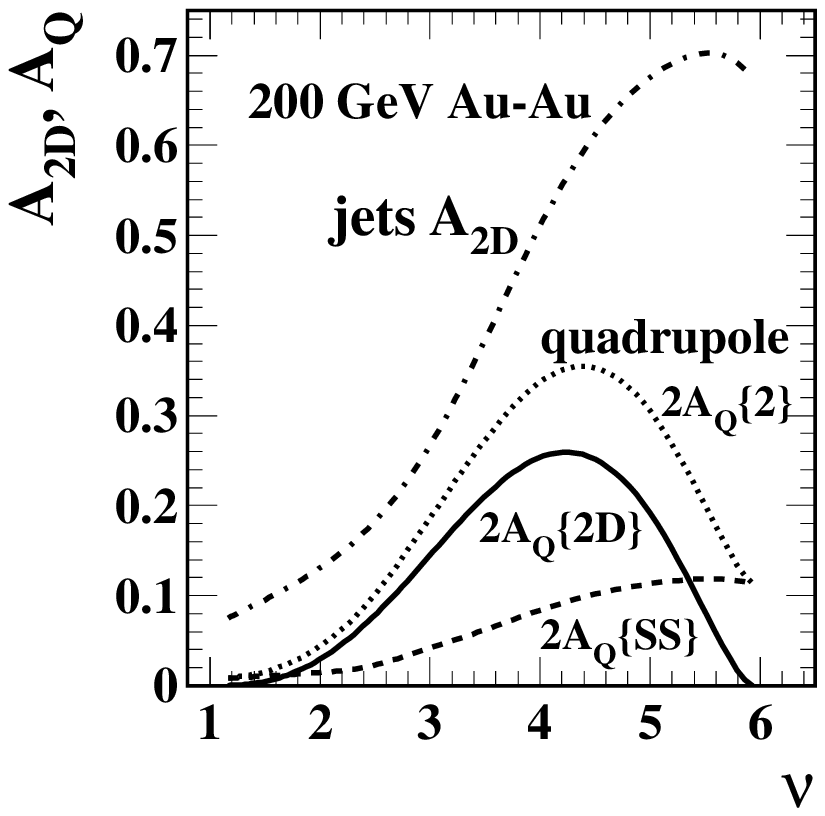}
  \put(-99,65) {\bf(c)}
\hfil
\caption{\label{tfig5}
(a) Universal quadrupole trends (solid, dashed curves) compared to published $v_2$ data (points) on $b$, (b) same on $N_{bin}$, (c) jet yield (dash-dotted curve) vs quadrupole components.
}  
 \end{figure}

Figure~\ref{tfig5} (a) (solid and dashed curves) shows a universal trend on centrality and collision energy inferred from 2D model fits, with $A_Q\{2D\}(b) = 0.0045 R(\sqrt{s_{NN}}) \epsilon_{opt}^2 N_{bin}$~\cite{davidhq}. The trends on $b/b_0$ are well described by Gaussians. Panel (b) shows the same relation in linear form on $R(\sqrt{s_{NN}}) N_{bin}$. Published $v_2\{EP\} \approx v2\{2\}$ data shown as points for comparison~\cite{2004} deviate strongly from the $\{2D\}$ trend. By combining results from Refs.~\cite{anomalous} and \cite{davidhq} we infer the relation $A_Q\{EP\} \approx A_Q\{2\} = A_Q\{2D\} + A_Q\{SS\}$ (dotted curve) as shown in panel (c). 2D model fits distinguish the jet-related quadrupole component of  the SS 2D peak $A_Q\{SS\}$ (``nonflow'') from the nonjet quadrupole $A_Q\{2D\}$. Such model fits are essential to disentangle jet-related from nonjet (possibly hydro-related) structure. The dotted curve in panel (c) is also plotted in panels (a) and (b) where it describes the published $v_2\{EP\}$ data (open squares) accurately.

\section{``Higher harmonics'' and \auau data}

The {\em jet-related} quadrupole and other jet-related multipoles (``higher harmonics'') can be inferred from SS 2D peak properties.

 \begin{figure}[h] \hfil
  \includegraphics[width=1.9in,height=1.6in]{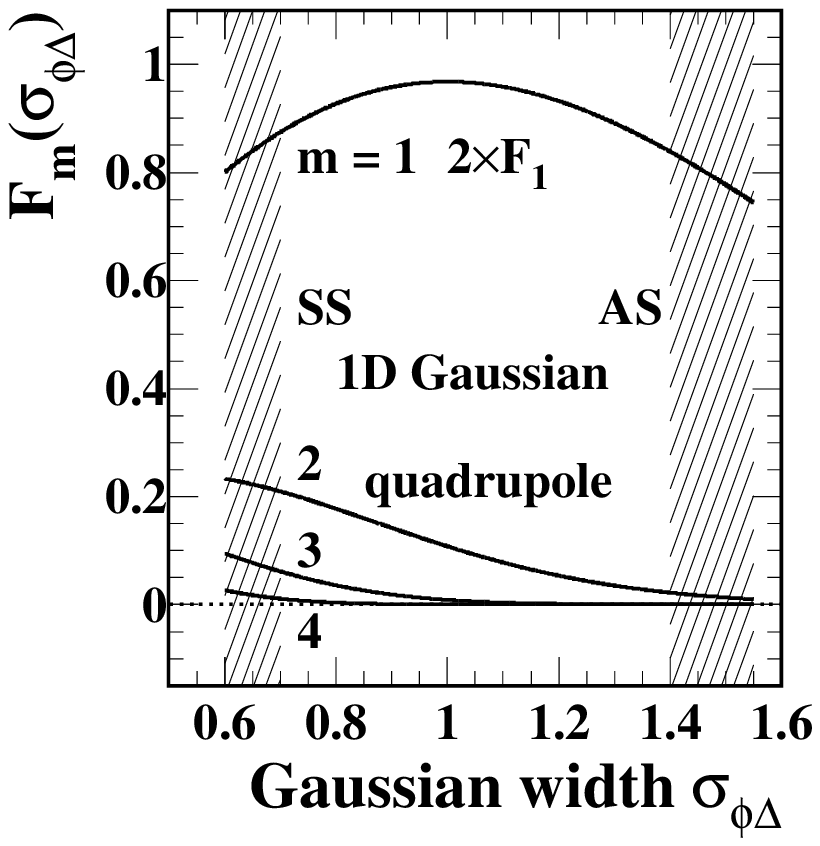} \hfil
  \put(-45,85) {\bf(a)}
  \includegraphics[width=1.9in,height=1.6in]{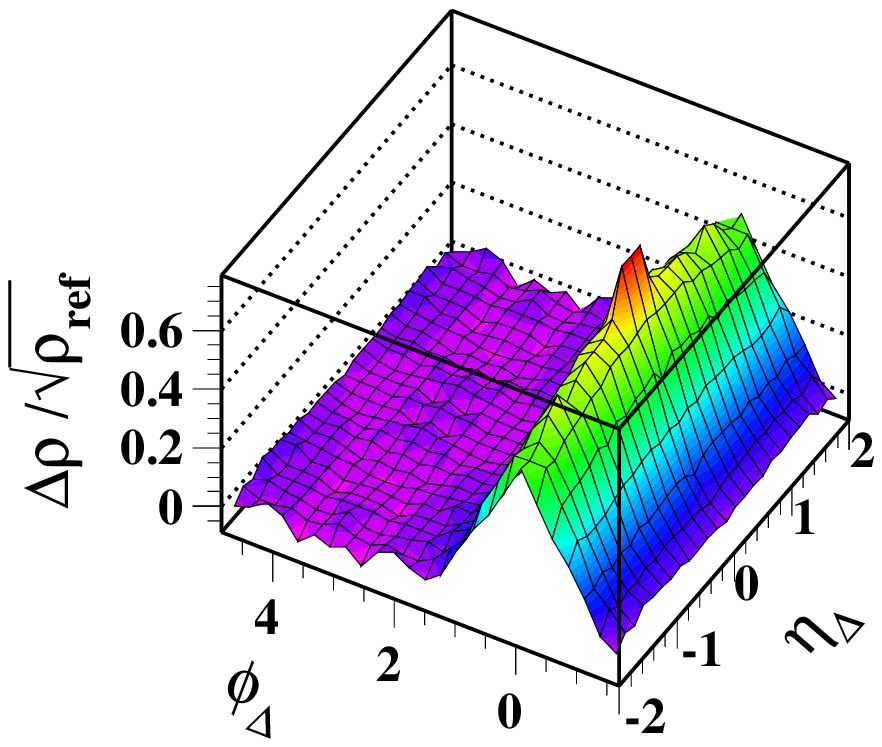} \hfil
   \put(-120,85) {\bf(b)}
 \includegraphics[width=1.9in,height=1.6in]{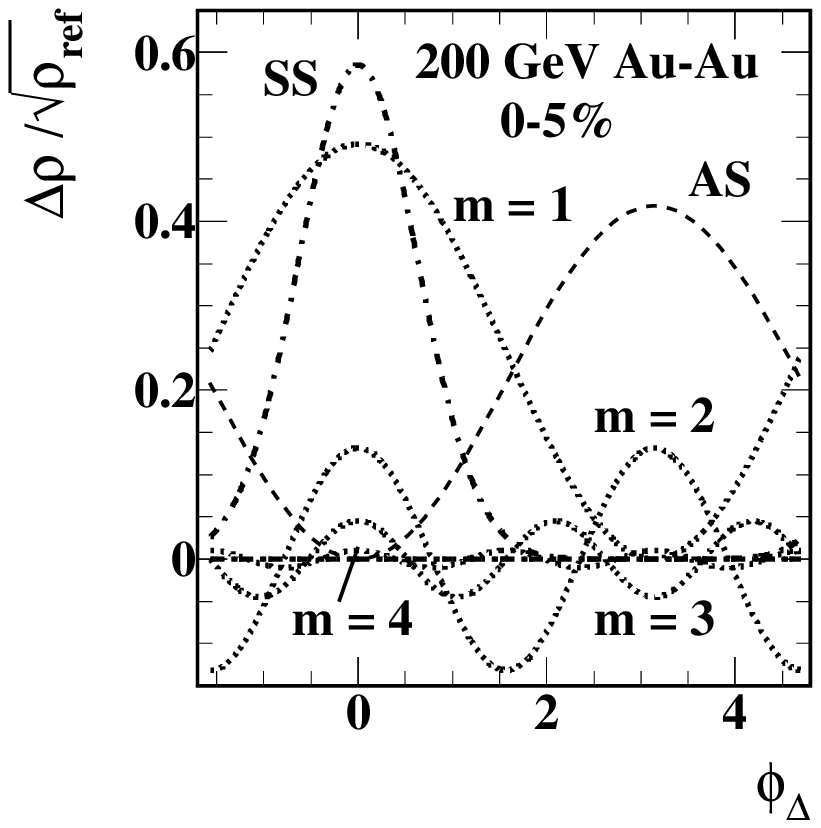}
  \put(-25,65) {\bf(c)}
\hfil
\caption{\label{tfig6}
(a) Fourier components of 1D Gaussian vs width, (b) SS 2D peak from 0-5\% central 200 GeV Au-Au collisions, (c), Azimuth multipoles for 1D Gaussian with width $\sigma_{\phi_\Delta} \approx 0.65$.
}  
 \end{figure}

Figure~\ref{tfig6} (a) shows the first few Fourier amplitudes for a {\em periodic} 1D Gaussian peak array on azimuth with width $\sigma_{\phi_\Delta}$~\cite{tzyam}. Typical widths for SS and AS peak arrays are indicated by the hatched regions. The amplitudes are given by $F_m(\sigma_{\phi_\Delta}) = \sqrt{2/\pi}\, \sigma_{\phi_\Delta} \exp(-m^2 \sigma_{\phi_\Delta}^2/2)$.
Whereas the broad AS 1D peak is well-described by a dipole alone the narrower SS 2D peak requires several multipoles. Panel (b) shows the 0-5\% central \auau data from Fig.~\ref{tfig1} (c) with the fitted AS dipole subtracted. All that remains is the SS 2D peak since $A_Q\{2D\}(b) \approx 0$ for that centrality~\cite{davidhq,davidhq2}. Multipole amplitudes are then given by $A_X\{SS\} \equiv \rho_0(b) v_m^2\{SS\}(b) = F_m(\sigma_{\phi_\Delta}) G(\sigma_{\eta_\Delta},\Delta \eta) A_{2D}$, where factor G represents projection of the SS 2D peak onto 1D azimuth, and subscript $X$ denotes multipoles $D,~Q,~S,~O$ for dipole, etc. Panel (c) shows the SS 1D peak (dash-dotted curve), the AS 1D peak (dashed curve) and several multipoles from the SS peak (dotted curves). For 0-5\% central \auau collisions the SS 2D peak is the {\em only} source of ``higher harmonics''~\cite{multipoles}.

\section{``Triangular flow''}

So-called ZYAM background subtraction in dihadron analysis leaves a double-peaked structure in the AS azimuth region that had been interpreted in terms of ``Mach cones''~\cite{tzyam}. The same structure has recently been reinterpreted in terms of ``triangular flow''~\cite{gunther}.


 \begin{figure}[h] \hfil
  \includegraphics[width=2.5in,height=2in]{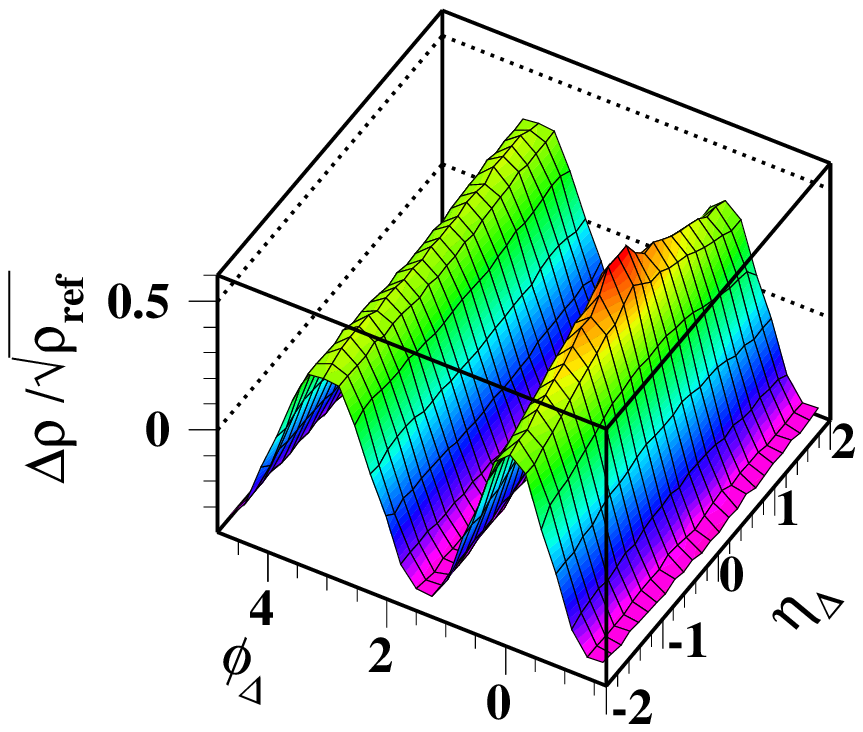} \hfil
  \includegraphics[width=3in,height=2.5in]{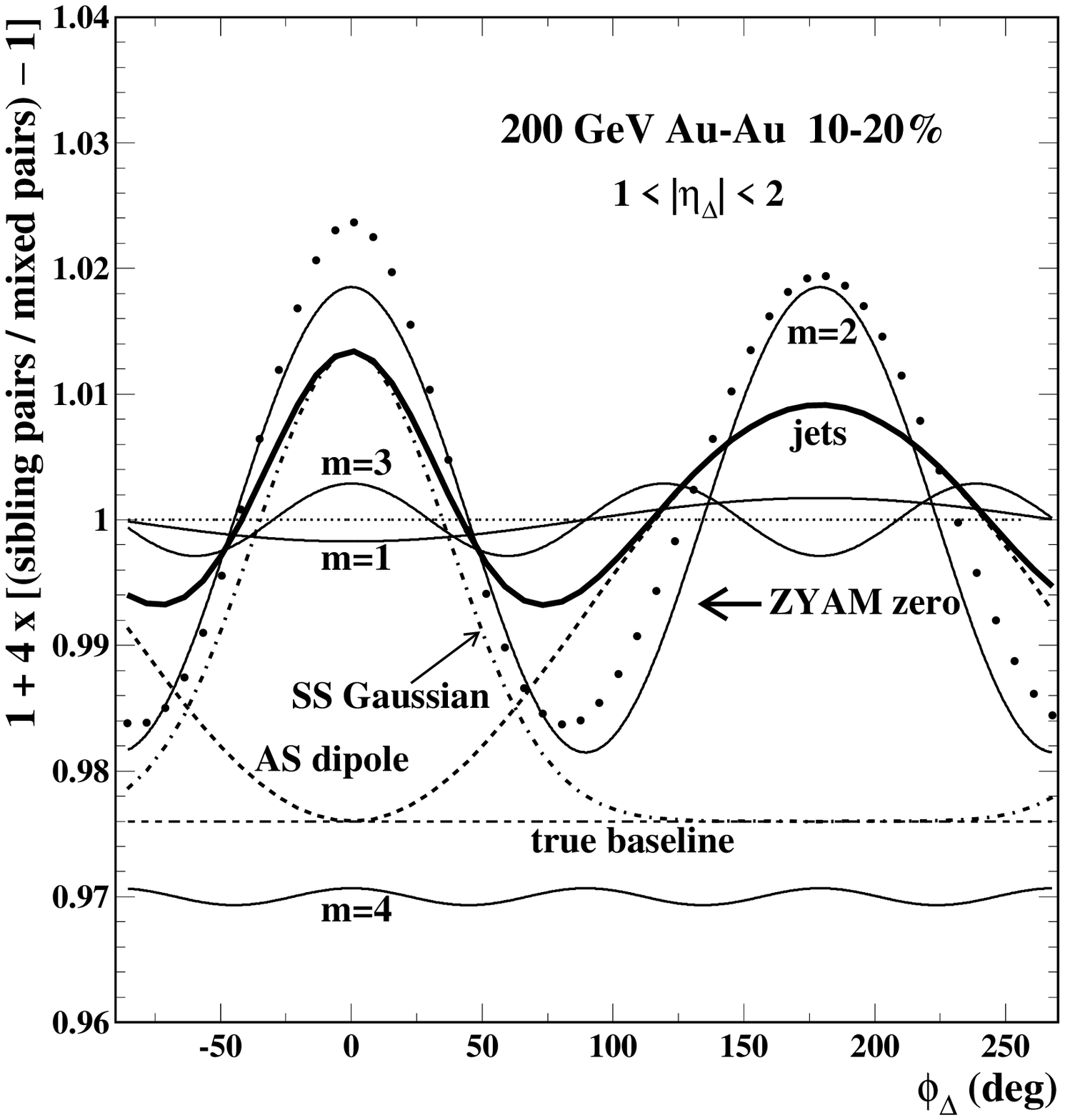}
\hfil
\caption{\label{tfig7}
Left: Angular correlation data from 200 GeV Au-Au collisions,
Right: 1D Fourier decomposition of similar data interpreted to reveal ``triangular flow'' ($m=3$ solid curve).
}  
 \end{figure}

Figure~\ref{tfig7} (left) shows 2D angular correlations from 10-20\% central 200 GeV \auau collisions. The unprojected 2D data are accurately described by a 2D model including SS 2D and AS 1D peaks and a nonjet quadrupole~\cite{anomalous}. Figure~\ref{tfig7} (right) shows a Fourier series decomposition (thin solid curves labeled $m = 1\ldots4$) of the 1D azimuth projection of equivalent data (points). The plot is based on Fig.~1 (right) of Ref.~\cite{gunther}. The 2D data have been projected from the interval $1 < |\eta_\Delta| < 2$ intended to exclude ``short-range'' jet structure. The left panel indicates that such exclusion cuts  are arbitrary and cannot distinguish jet-related from nonjet structure in more-central \auau collisions. The jet-related SS 1D (dash-dotted) and AS 1D (dashed) curves derived from 2D model fits sum to the bold solid curve representing projection of the data histogram in the left panel minus the fitted nonjet quadrupole. 

Some Fourier terms represent multiple correlation features:   $v_1^2\{2\} = v_1^2\{SS\} - v_1^2\{AS\}$, $v_2^2\{2\} = v_2^2\{2D\} + v_2^2\{SS\}$, $v_3^2\{2\} = v_3^2\{SS\}$ and $v_4^2\{2\} = v_4^2\{SS\}$ for $m=1\ldots4$ respectively~\cite{tzyam}.  The apparently small $m=1$ dipole term ({\em not} ``directed flow'') has been described as  ``global momentum conservation'' but is actually the difference between two large amplitudes  representing parton scattering to jets. The $m=3$ sextupole component is denoted ``triangular flow.'' In effect, the jet-related SS 2D peak is ``fragmented'' to nearly cancel the AS 1D peak and provide large ``nonflow'' contributions to $v_2^2\{2\}$ and the ``higher harmonic flows.''


The centrality dependence of ``triangular flow'' and other ``higher harmonics'' singly and in ratios has been compared to hydrodynamic calculations.

 \begin{figure}[h] \hfil
  \includegraphics[width=1.9in,height=1.6in]{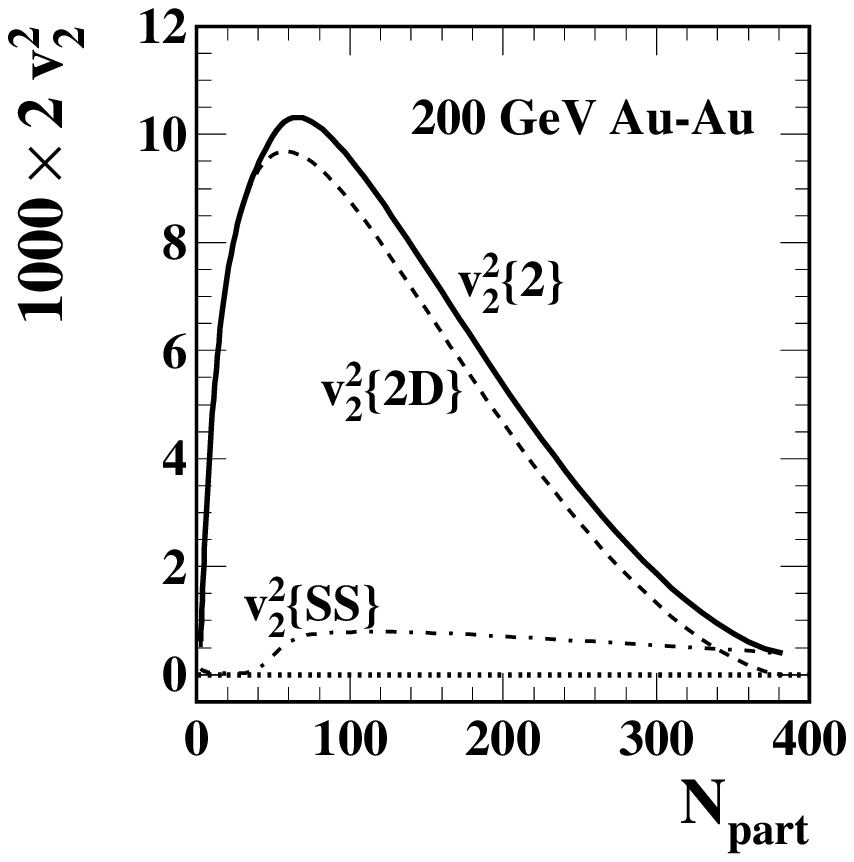} \hfil
  \put(-100,75) {\bf(a)}
   \includegraphics[width=1.9in,height=1.6in]{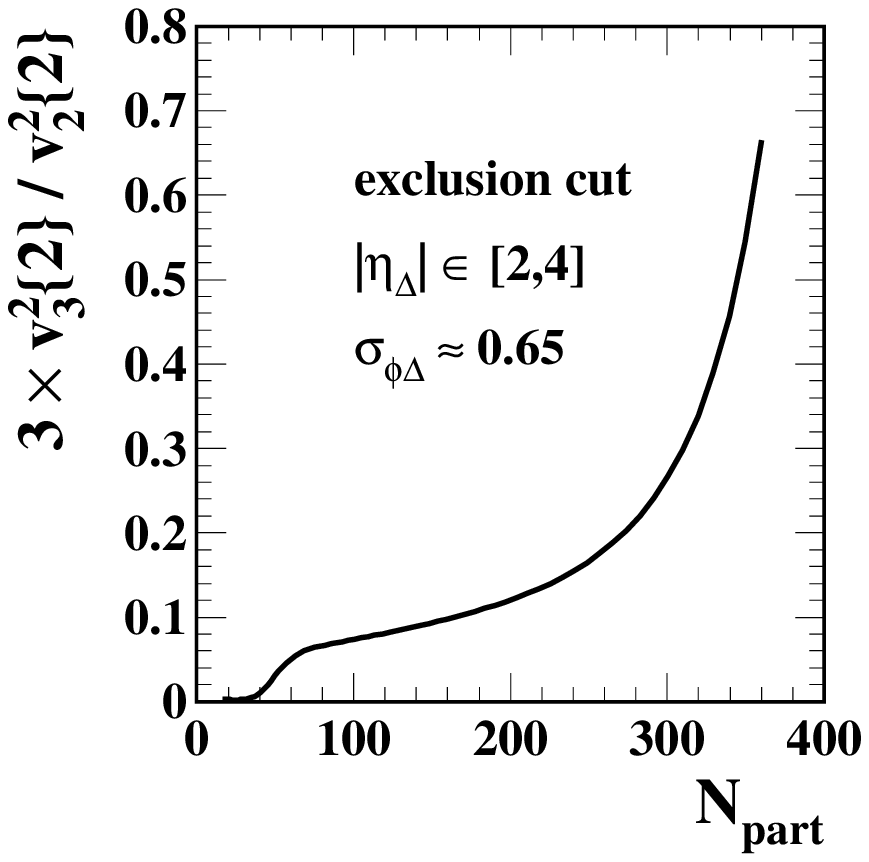} \hfil
  \put(-100,75) {\bf(b)}
  \includegraphics[width=1.9in,height=1.6in]{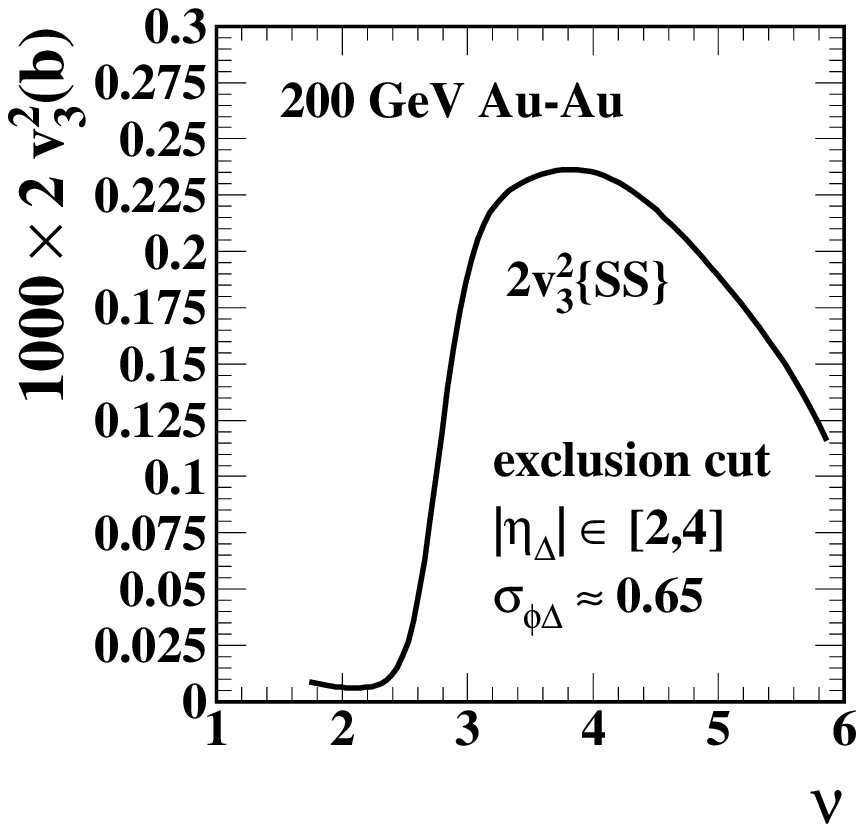} \hfil
\hfil
  \put(-112,75) {\bf(c)}
\caption{\label{tfig8}
(a) Jet-related $\{SS\}$, nonjet $\{2D\}$ and total $\{2\}$ azimuth quadrupole components vs $N_{part}$, (b) Total sextupole/quadrupole ratio vs $N_{part}$, (c) Jet-related sextupole component vs $\nu$.
}  
 \end{figure}

Figure~\ref{tfig8} (a) shows quadrupole amplitudes $v_2^2$ plotted vs participant number $N_{part}$ which obscures the important lower half of the fractional cross section. $v_2^2\{2\}$ represents the quadrupole amplitude for nonjet {\em and} jet-related two-particle correlations. $v_2^2\{SS\}$ denotes the quadrupole component of the jet-related SS 2D peak. The data have been projected from within $2 <|\eta_\Delta| < 4$ assumed to exclude jet structure. However, for more-central \auau collisions (above the sharp transition, $N_{part} > 50$, $\sigma/\sigma_0 < 0.5$ or $\nu > 3$) that strategy is ineffective. 

In panel (b) the ratio $v_3^2 / v_2^2$ is easily explained~\cite{multipoles}. The nonjet~\cite{davidhq} and jet-related (``nonflow'')~\cite{davidhq2} quadrupole trends are accurately known and result in the solid curve. 
In more-peripheral collisions the nonjet quadrupole dominates $m = 2$. In more-central collisions the nonjet quadrupole goes to zero and the $v_3^2/v_2^2$ ratio goes asymptotically to $F_3 / F_2 \approx 1/3$, so the plotted ratio goes to 1. Figure~\ref{tfig8} (c) shows $v_3^2\{SS\}(b)$ derived from the analysis in Ref.~\cite{anomalous} plotted on $\nu$. The effect of the $\eta$-exclusion cut is clear: As the SS 2D peak elongates on $\eta$ above the ST at $\nu = 3$ it extends into the cut acceptance. The sextupole component of the jet-related SS 2D peak accounts for all centrality dependence of ``triangular flow.''


More recently, attempts to determine the $\eta$ dependence of ``triangular flow'' have emerged.

 \begin{figure}[h] \hfil
  \includegraphics[width=1.9in,height=1.6in]{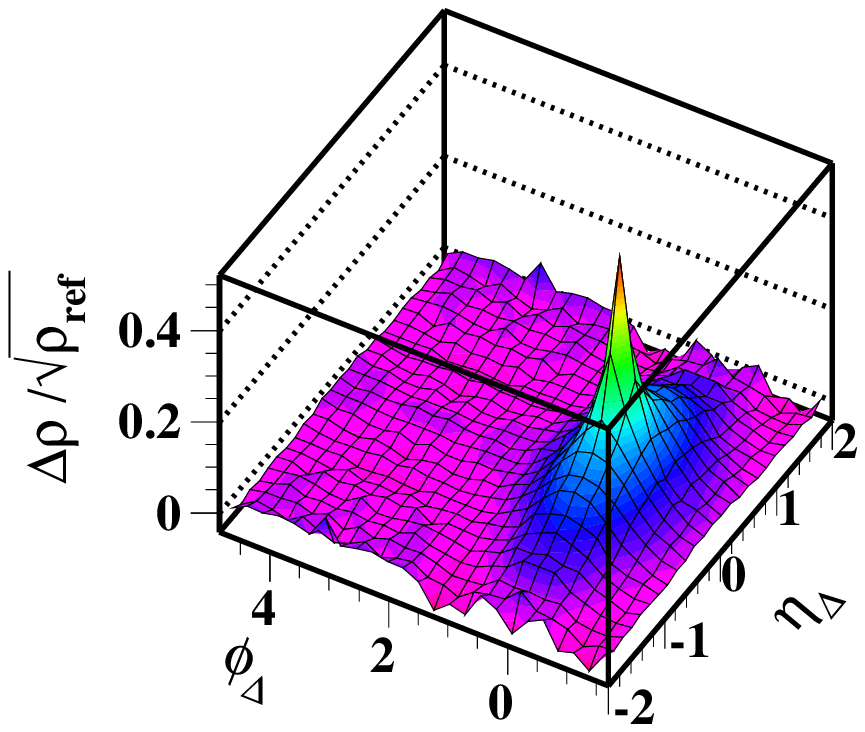} \hfil
   \put(-120,85) {\bf(a)}
  \includegraphics[width=1.9in,height=1.6in]{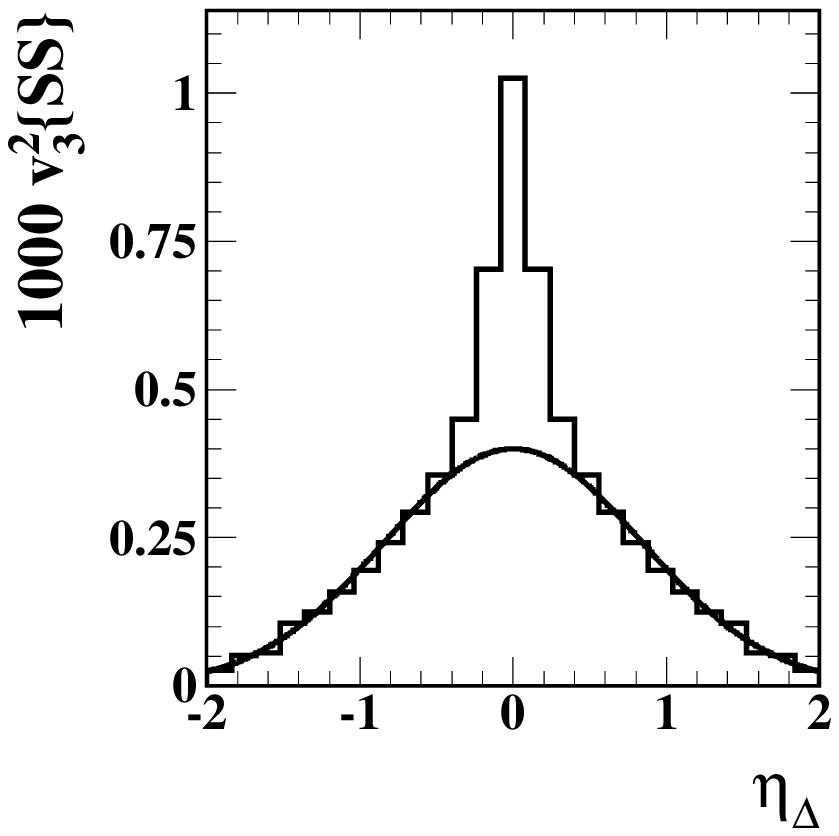} \hfil
  \put(-95,55) {\bf(b)}
  \includegraphics[width=1.9in,height=1.6in]{mike-final200-30aslopex} \hfil
  \put(-95,55) {\bf(c)}
\caption{\label{tfig9}
(a) 2D angular correlations from 45-55\% central 200 GeV Au-Au collisions, (b) Corresponding sextupole component vs $\eta_\Delta$, (c) SS 2D peak amplitude vs mean path length $\nu$.
}  
 \end{figure}

Figure~\ref{tfig9} (a) shows 2D angular correlations from 45-55\% central 200 GeV Au-Au collisions~\cite{anomalous}. The fitted AS dipole and nonjet quadrupole terms have been subtracted leaving a broader jet-related SS 2D peak and a superposed narrow peak from Bose-Einstein correlations and conversion-electron pairs. The properties of the broader SS peak are completely consistent with pQCD jet expectations (amplitude, charge composition, $p_t$ structure).

Figure~\ref{tfig9} (b) shows the result of a Fourier series analysis of data from each $\eta_\Delta$ bin of panel (a) to obtain the $m=3$ sextupole component (histogram) compared to that for the fitted 2D Gaussian from Ref.~\cite{anomalous} (solid curve). The SS 2D peak does have a sextupole component (among other multipoles), but that does not imply the peak structure represents ``triangular flow.''

Figure~\ref{tfig9} (c) repeats the SS 2D peak amplitude vs centrality. The data in panel (a) correspond to $\nu \approx 2.8$---{\em below the sharp transition} where Glauber linear superposition (\auau transparency) is observed. There is no reason to expect any collective behavior in that centrality interval. By misapplication of Fourier analysis jet-related angular correlations become ``triangular flow.''

\section{``Higher harmonic flows'' at the LHC}

Recently, a trend has emerged to attribute all angular correlation structure projected onto 1D azimuth to ``higher harmonic flows''~\cite{luzum}.

 \begin{figure}[h] \hfil
  \includegraphics[width=3in,height=3in]{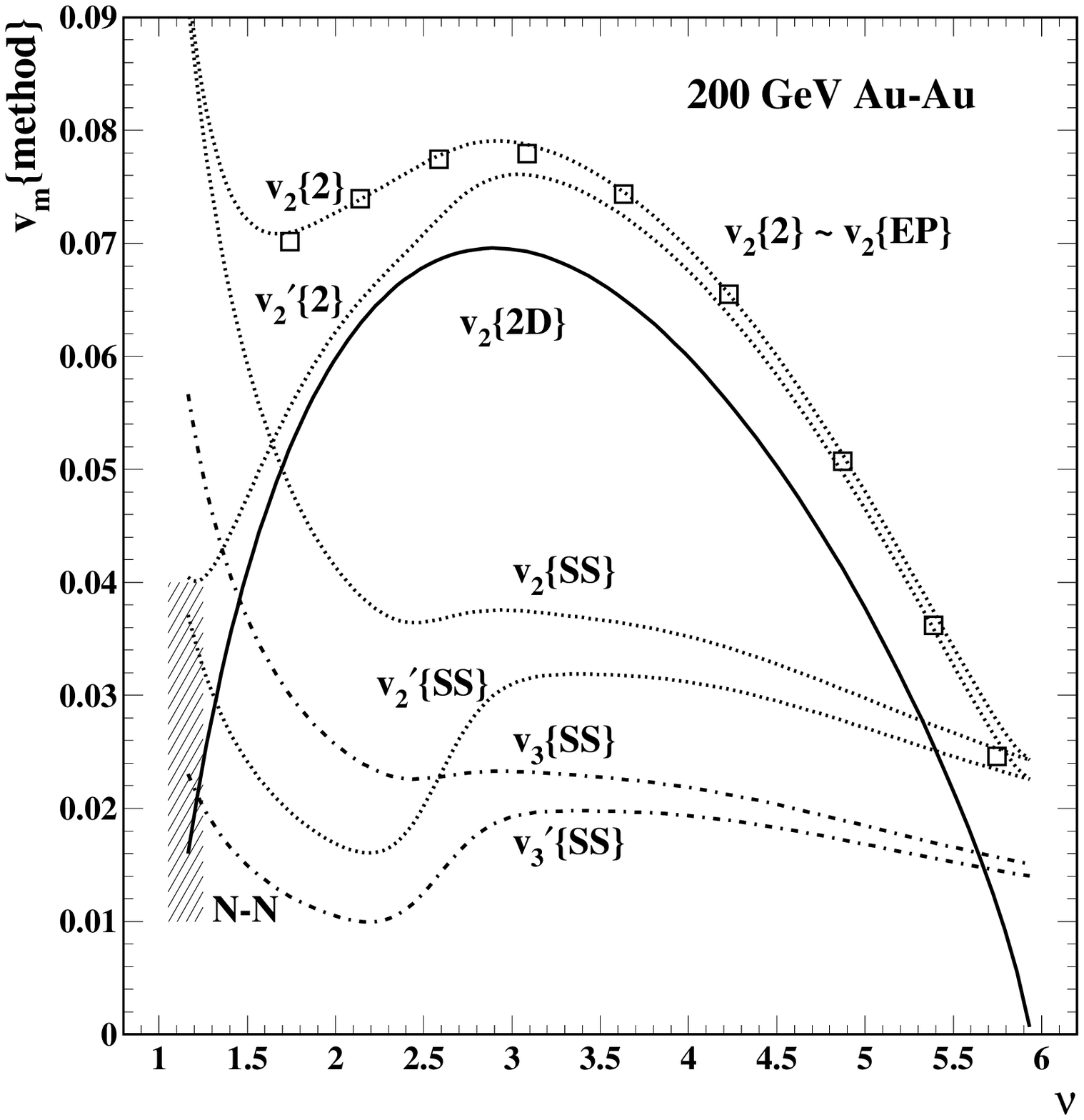} \hfil
   \includegraphics[width=3in,height=3in]{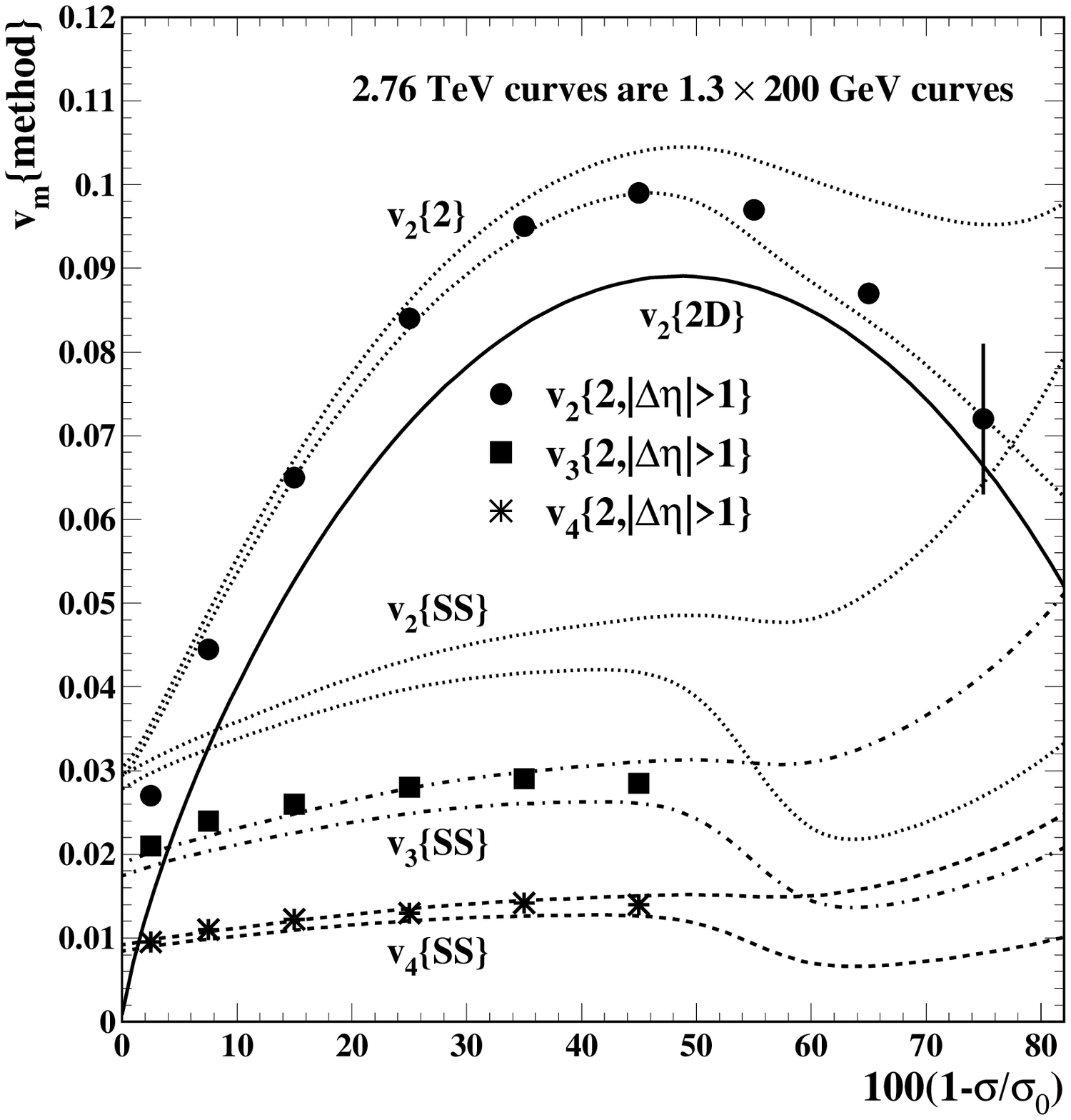} \hfil
\caption{\label{tfig10}
Left: Jet-related $\{SS\}$ multipoles for the SS 2D peak from 200 GeV Au-Au collisions,
Right: Curves from the left panel  compared to 2.76 TeV \pbpb data (points).
}  
 \end{figure}

Figure~\ref{tfig10} (left) shows Fourier amplitudes for 2D angular correlation structure from 200 GeV \auau collisions~\cite{anomalous} projected onto 1D azimuth from within the interval $1 < |\eta_\Delta| < 2$, an {\em $\eta$ exclusion cut} expected to exclude jet structure~\cite{multipoles}. As usual, $v_m\{SS\}$ represent Fourier amplitudes from the SS 2D peak,  $v_2\{2D\}$ represents the nonjet quadrupole derived from 2D model fits and $\{2\} \approx \{EP\}$ represent multipoles derived from all angular correlations ($\{EP\}$ denotes the so-called ``event-plane'' $v_2$ method~\cite{quadmeth}). Primes indicate that an $\eta$ exclusion cut has been imposed. Unprimed quantities indicate no exclusion cut. Those trends provide predictions for any results from analysis directed toward ``higher harmonic flows.''

Figure~\ref{tfig10} (right) shows analysis results (points) from 2.76 TeV \pbpb collisions~\cite{alice}. The curves are the trends from the left panel multiplied by common factor 1.3 and plotted on $1 - \sigma / \sigma_0$ instead of $\nu$. The curves describe the LHC data well, especially the effects of the $\eta$ exclusion cuts for the $v_2\{2\}$ data, suggesting that the sharp transition in jet-related structure persists at LHC energies. The nonjet quadrupole centrality trend also seems similar to that at 200 GeV. It is unfortunate that the $v_3$ and $v_4$ data do not extend below the sharp transition near $\sigma / \sigma_0 = 0.5$ ($\nu \approx 3$). The factor 1.3 has been attributed to a change in the $p_t$ spectrum at the higher energy~\cite{alicept}, presumably due to a larger jet contribution.

\section{Summary}

$p_t$-integral 2D angular correlations within a limited $\eta$ acceptance from 200 GeV \auau collisions include three main features: a monolithic SS 2D peak, an AS 1D peak and a quadrupole uniform on $\eta$ and distinct from the SS and AS peaks. For all \auau centralities most properties of the SS 2D peak and AS 1D peak have been linked quantitatively to jet-related pQCD predictions. At 200 GeV the principal deviations from in-vacuum jet phenomenology in more-central collisions are (a) elongation of the SS 2D peak on $\eta$ and (b) {\em increase} of the SS and AS peak amplitudes beyond the \nn binary-collision scaling expected for in-vacuum jets. Modification (b) is quantitatively consistent with pQCD-described alteration of parton fragmentation that conserves jet energy.

The persistence of a relatively large jet-correlated hadron yield in more-central \auau collisions (30\% of the final state) is inconsistent with interpretation of RHIC collisions in terms of a thermalized, flowing, dense partonic medium, a so-called ``perfect liquid.'' The response has consisted of attempts to reinterpret jet-related structures, especially the elongated SS peak renamed the ``soft ridge,'' as flow phenomena. Typical strategies reduce to projecting all 2D angular correlations within some $\eta$ interval onto 1D $\phi$ and fitting the result with a single sinusoid or a Fourier series with several terms. Each Fourier coefficient is interpreted as a flow.

Such procedures can be questioned for at least three reasons: (a) Substantial information in 2D angular correlations that may falsify flow interpretations is discarded in the projection to 1D azimuth, (b) detailed properties of nominally jet-related structure for all \auau centralities correspond to pQCD jet expectations, including $p_t$ structure, charge structure and centrality variation and (c) a Fourier series by definition can describe any periodic distribution on 1D azimuth, but interpretation of Fourier coefficients in terms of flows is {\em not necessary}. Other interpretations may be (and are) more likely within the context of high-energy nuclear collisions.

The systematics of inferred ``higher harmonic flows'' are {\em more efficiently represented} by the measured properties of the SS 2D peak. Typical 1D Fourier representations cannot describe the unprojected 2D angular correlations---the 1D azimuth model is falsified by the 2D data. The SS 2D peak biases all inferred $v_m\{2\}$ data as ``nonflow'' in known ways. The SS peak is the only source of ``higher harmonic flows'' in essentially all cases. The nonjet quadrupole, isolated from SS and AS peak structure by 2D model fits, has systematics very different from nominally jet-related structure and is directly connected to the simplest (optical) representation of initial-state \aa eccentricity.
A substantial body of evidence supports the conclusion that the so-called ``soft ridge'' or same-side 2D peak is a jet phenomenon even in more-central \auau collisions, albeit the fragmentation process is modified in a manner described for the most part by pQCD. 

\ack
This work was supported in part by the Office of Science of the U.S. DOE under grant DE-FG03-97ER41020.


\end{document}